\documentclass{elsart1p}

\newcommand{\beq}{\begin{equation}}
\newcommand{\eeq}{\end{equation}}
\newcommand{\ba}{\begin{eqnarray}}
\newcommand{\ea}{\end{eqnarray}}

\newcommand{\sig}{\langle\sigma\rangle}
\newcommand{\et}{\langle\eta\rangle}
\newcommand{\al}{\vec\alpha_0}

 \usepackage{graphicx}
 \usepackage{epsfig}

\usepackage{amssymb}
\usepackage{rotating}

\begin{document}

\begin{frontmatter}



\title{CP Violation in the Linear Sigma Model}


\author{Ana J\'ulia {\sc Mizher} and
Eduardo S. {\sc Fraga}}

\address{Instituto de F\'\i sica, Universidade Federal do Rio de Janeiro, \\
Caixa Postal 68528, Rio de Janeiro, RJ 21941-972, Brazil}

\begin{abstract}
Motivated by the possibility of the formation of CP-odd domains in heavy ion collisions, we investigate the effects of CP violation on the chiral transition within the linear sigma model with two flavors of quarks. We also study how the CP-odd system is affected by the presence of a strong magnetic field, that is presumably generated in a non-central heavy ion collision. We find that both ingredients play an important role, influencing drastically the nature of the phase transition and the critical temperature.

\end{abstract}

\end{frontmatter}

\section{Introduction}

In Quantum Chromodinamics (QCD), non-trivial gauge field configurations produce a term in the Lagrangean that violates the CP symmetry. So far experiments have not detected this symmetry breaking, indicating that the overall coefficient of this term, if nonvanishing, has a very small magnitude. It is known  that the CP symmetry can not be spontaneously broken in the vacuum of QCD \cite{vafa_witten}. However, this statement is not valid for thermal QCD \cite{cohen}. In this vein, Kharzeev, Pisarski and Tytgat \cite{kharzeev1} proposed ten years ago that, during the chiral phase transition, the high temperature matter produced in a heavy ion collision might reach a metastable state. This state could be described by taking into account the topological term mentioned above, and would decay via processes that violate CP. This scenario was explored in detail in Refs. \cite{kharzeev2, kharzeev3}. Later, suitable experimental measurements \cite{kharzeev4} and observables \cite{voloshin} were proposed. The experimental signatures are based on a mechanism of charge  separation after a heavy ion collision. To observe the effect, it is essential that the collisions are non-central, in order to generate a strong magnetic field that is responsible for the current that will result in charge separation at the end. Estimates of the magnitude of the magnetic field generated in heavy ion collisions at RHIC and the LHC indicate that they are strong enough to provide the necessary conditions for the effect to be measurable \cite{kharzeev2}.

In this work we study how the ingredients mentioned above affect the thermally-induced chiral transition in QCD. We investigate the influence of CP violation and a strong magnetic field on a system described by an effective theory containing quarks and mesons as degrees of freedom.  For this purpose, we first analyze how the chiral transition is modified by the presence of a topological term that mimics the action of non-trivial gauge field configurations, responsible for the existence of the CP-odd term in the action. Afterwards, we study the changes brought about by the presence of a strong magnetic background. As an effective theory, we adopt the linear sigma model (LSM), defined by the following Lagrangean:
\ba
\nonumber
\mathcal{L}&=& \frac{1}{2} Tr(\partial_\mu\phi^\dagger \partial^\mu \phi) + \frac{a}{2} Tr(\phi^\dagger \phi) - \frac{\lambda_1}{4} [Tr(\phi^\dagger \phi)]^2
-\frac{\lambda_2}{4} Tr[(\phi^\dagger \phi)^2]\\
&&+ \frac{c}{2}[e^{i\theta}det(\phi) + e^{-i\theta}det(\phi^\dagger)] + Tr[h(\phi +\phi^\dagger)] \; ,
\label{Lini}
\ea
where $\phi = \frac{1}{\sqrt{2}}(\sigma + i\eta) + \frac{1}{\sqrt{2}} (\vec \alpha + i \vec{\pi}) \cdot \vec{\tau}$. Here, CP violation is encoded in the determinant term, known as the 't Hooft determinant term, whose magnitude is quantified by the parameter $\theta$.
In our treatment, quarks constitute a thermalized fluid that provides a background in which the long-wavelength field of the chiral condensate evolves. In this context, free quarks  are absent at zero temperature, becoming relevant degrees of freedom only at finite temperature, where we include quark thermal fluctuations to one loop. We couple mesons to quarks via a Yukawa coupling, and the Lagrangean, including quarks, becomes:
\ba
\nonumber
 \mathcal{L} &=& \frac{1}{2}(\partial_\mu\sigma)^2 +\frac{1}{2}(\partial_\mu\vec\pi)^2 + \frac{1}{2}(\partial_\mu\eta)^2 + \frac{1}{2}(\partial_\mu\al)^2
- V(\sigma,\eta,\vec\pi,\al) \\
&&+\bar\psi(i\gamma^\mu\partial_\mu)\psi -g\bar\psi(-i\gamma^5\vec\tau\cdot\vec\pi +\sigma)\psi
 - g\bar\psi(\gamma^5\vec\tau\cdot\al - i\gamma^5\eta)\psi \; ,
\ea
where $V$ is the mesonic self-interaction potential that can be easily extracted from (\ref{Lini}).

In a mean field analysis, we take $\sigma = \sig +
\sigma'$ and $\eta = \et + \eta'$, and assume that the remaining condensates vanish.
The effective mass for the quarks is obtained from the Yukawa
coupling within this approximation, yielding $M=g\sqrt{\sig^2 + \et^2}$.

The potential is composed by a classical piece of the form
\ba
\nonumber
V_{cl} &=& \left(\frac{-a - c \cos\theta}{2}\right)\sig^2 - H\sig
+ c \sin\theta \sig\et \nonumber\\
&&+\left(\frac{-a+c \cos\theta}{2}\right) \et^2
+\frac{1}{4} \left(\lambda_1 + \frac{\lambda_2}{2}\right) (\sig^2 +\et^2)^2
\label{eq:eff_pot}
\ea
where $H=\sqrt{2} h$, and fluctuations. The parameters in the lagrangian are fixed to reproduce
meson masses and the pion decay constant in the vacuum.

\begin{figure}[htbp]
\begin{center}
\includegraphics[width=4cm]{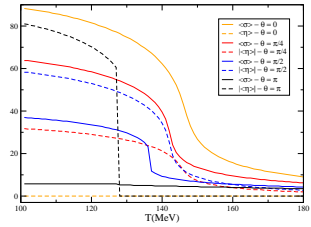}
\caption{Value of the condensates as functions of the temperature.}
\label{fig:minima}
\end{center}
\end{figure}

Including thermal corrections, we compute the effective potential as a function of $\sig$, $\et$, $\theta$, and the temperature $T$. For $\theta=0$, one recovers the expected behavior of the usual LSM. Keeping the temperature at zero and varying $\theta$, the minima of the effective potential rotate, though the local minimum never becomes the global minimum. Although a large value for the parameter $\theta$ in an effective model for QCD in the vacuum is not realistic, it illustrates some important features of the dependence of the effective potential on $\theta$. For $\theta=\pi$, a theoretically relevant case \cite{Boer:2008ct}, the minima are almost along the $\eta$ axis (never reaching the axis because of the explicit symmetry breaking built in our model).

As we increase the temperature, both condensates tend to melt. The $\eta$ condensate vanishes before, so that we find three different phases: one with $\sigma$ and $\eta$ nonzero condensates, another with only the $\eta$ condensate, and finally one where both condensates vanish. In Fig. \ref{fig:minima} one can see the condensates dropping to zero as the temperature is increased for different values of $\theta$. We can identify the expected crossover for $\sigma$ in this case, whereas for $\eta$ we find a first-order transition as one approaches $\theta=\pi$.

To investigate how this situation is modified by the presence of a strong magnetic background, we follow Ref. \cite{Fraga:2008qn}. Assuming a constant and homogeneous magnetic field $B$, we incorporate its effects by redefining the dispersion relations:
\begin{eqnarray}
p_{0n}^2=p_z^2+m^2+(2n+1)|q|B \quad , \quad p_{0n}^2=p_z^2+m^2+(2n+1-\sigma)|q|B \ ,
\end{eqnarray}
for scalars and fermions, respectively, $n$ being an integer, $q$ the electric charge,
and $\sigma$ the sign of the spin. In our effective model, the vacuum contribution to the potential will be modified by the magnetic field
through the coupling of the field to charged pions. For the thermal corrections, initially provided by pions and quarks, only part of the quark contribution is not exponentially suppressed. For the explicit expressions of these quantities see Ref. \cite{Fraga:2008qn}.

\begin{figure}[!ht]
\begin{center}
\begin{tabular}{cc}
 \epsfig{file=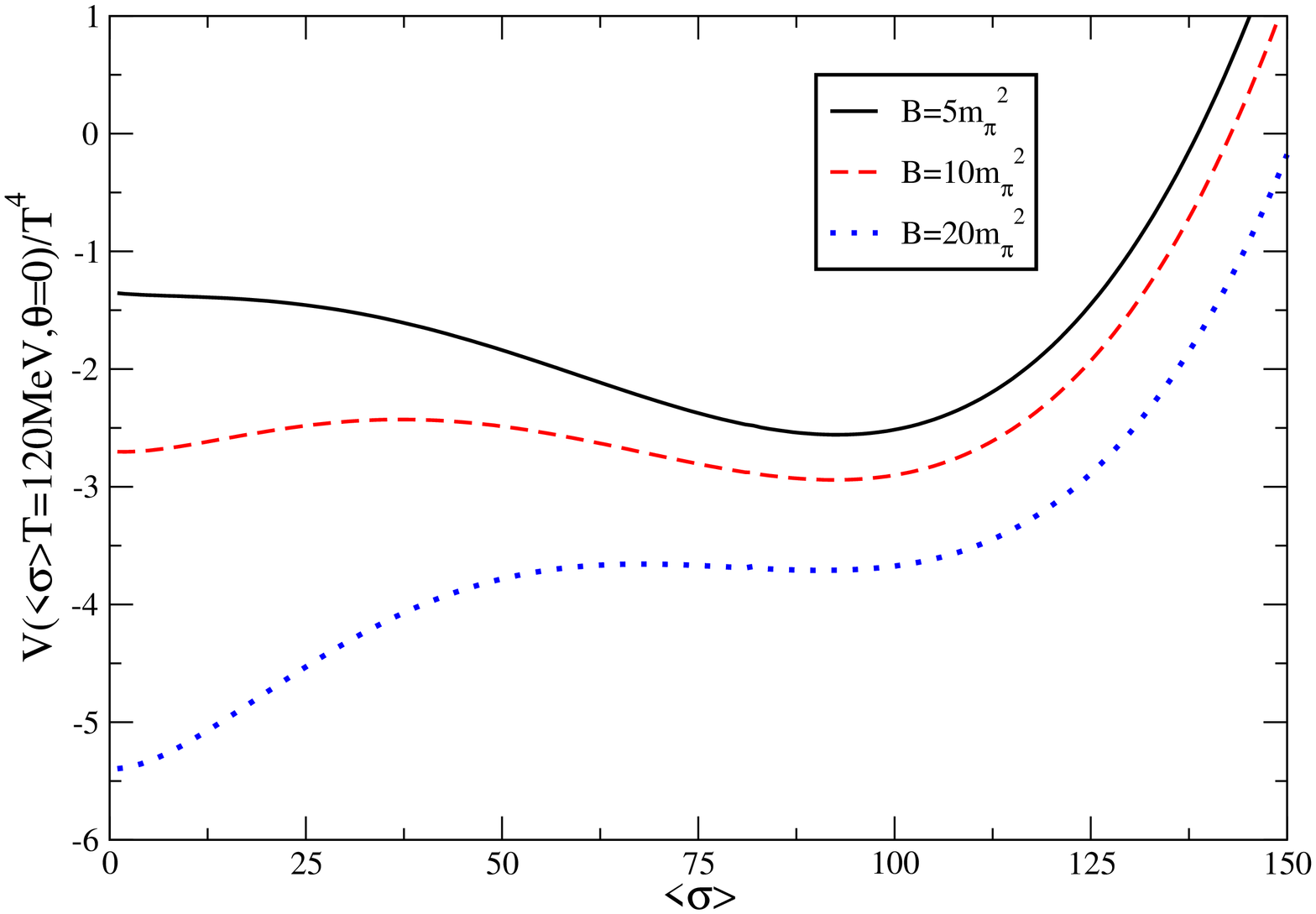,width=5.0cm,height=180pt,angle=270}&
 \hspace{0.5cm} \epsfig{file=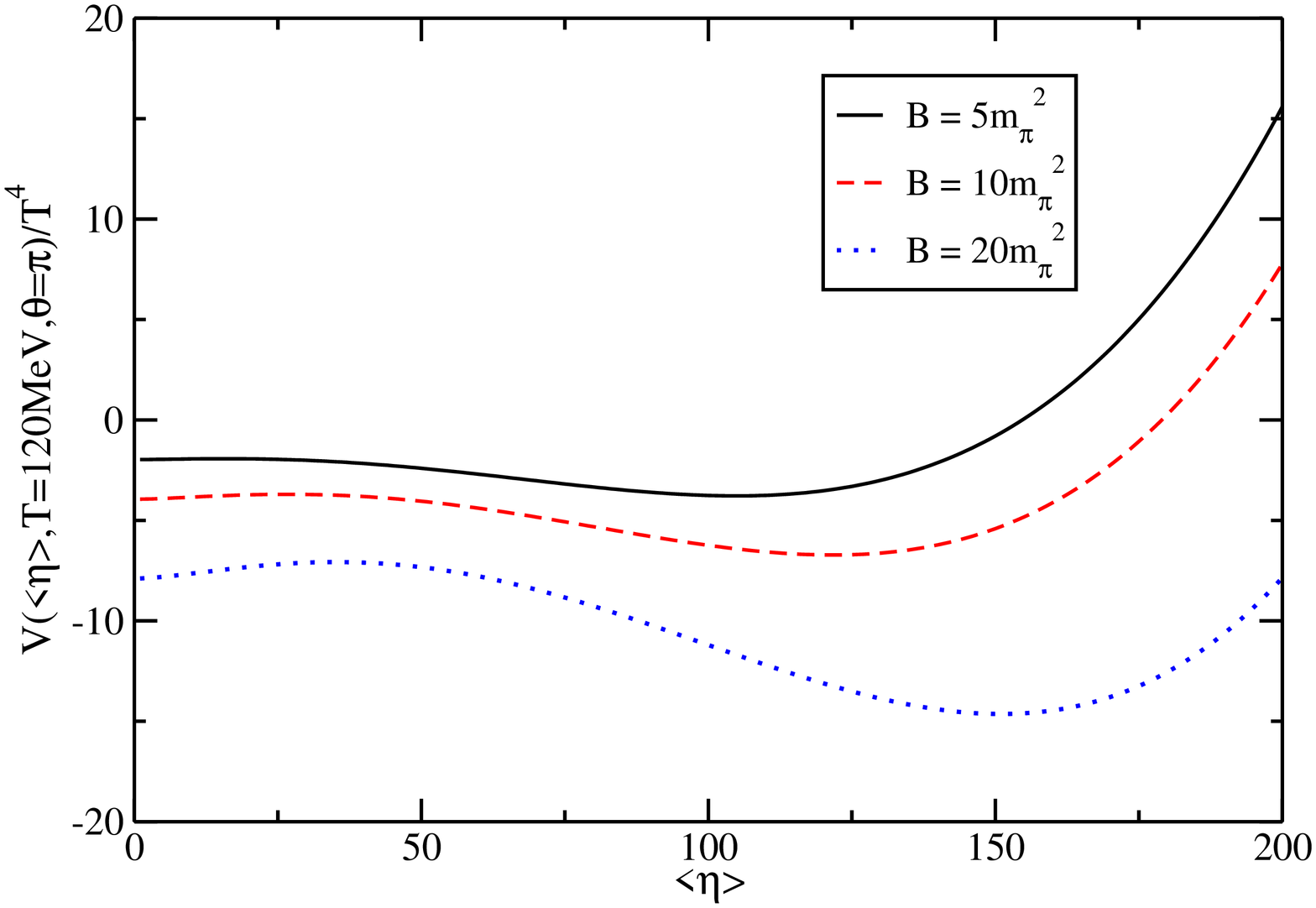,width=5.0cm,height=180pt,angle=270}\\
 (a) & \ \ \ \ \ \ (b)\\
\end{tabular}
\end{center}
\caption{Effective potential for different values of the magnetic field $B$. a) $\theta=0$ in the $\sigma$ direction \ \ \ b) $\theta=\pi$ in the $\eta$ direction.}
 \label{fig:several_B}
\end{figure}

Plots of the full effective potential, including effects from the magnetic field, are shown in Fig. \ref{fig:several_B}, which illustrates how dramatic the modifications in the behavior of the phase transition are, as expected from the results of \cite{Fraga:2008qn}. Increasing the magnitude of the field brings two main consequences: the nature of the phase transition is turned from a crossover to a first-order phase transition, and the critical temperature changes. In Fig. \ref{fig:several_B}, we plot the effective potential for different magnitudes of the field $B$ and same temperature, $T=120~$MeV. In (a) we plot the potential for $\theta=0$ in the $\sigma$ direction, and we can clearly see that the phase transition turns from a crossover to a first-order transition. The critical temperature is initially enlarged by the presence of the field but, from a critical value on, it starts to drop. For $\theta=\pi$, and in the $\eta$ direction, the presence of the field makes the first-order transition weaker. However, increasing the field the strength of the first-order transition is also increased.

CP violation and the presence of a strong magnetic background affect considerably the physical picture of the chiral
phase transition. Qualitative changes in the nature of the phase transition and quantitative modifications of the
critical temperature have large influence on the dynamics of the system, especially on the relevant time scales.
The dependence of the condensates on $\theta$ and $T$ is clearly non-trivial and must be explored on detail.
Results in this direction will be presented soon \cite{future}.

We thank D. Boer and J. K. Boomsma for fruitful discussions.

This work was partially supported by CAPES, CNPq, FAPERJ and FUJB/UFRJ.

\label{}




\begin{thebibliography}{00}

\bibitem{kharzeev1}D. Kharzeev, R.D. Pisarski, M.H.G. Tytgat,
Phys.Rev.Lett. {\bf 81}, \\512 (1998).


\bibitem{vafa_witten}
  C.~Vafa and E.~Witten,
  Phys.\ Rev.\ Lett.\  {\bf 53}, 535 (1984);
  C.~Vafa and E.~Witten,
  Nucl.\ Phys.\  B {\bf 234}, 173 (1984).



\bibitem{cohen}  S.~Bronoff and C.~P.~Korthals Altes,
  Phys.\ Lett.\  B {\bf 448}, 85 (1999)
  V.~Azcoiti and A.~Galante,
  Phys.\ Rev.\ Lett.\  {\bf 83}, 1518 (1999);
T.D. Cohen,
Phys.Rev.D {\bf 64}, 047704 (2001);
  X.~d.~Ji,
   ``Validity of the Vafa-Witten proof on absence of spontaneous parity
  Phys.\ Lett.\  B {\bf 554}, 33 (2003);
  M.~B.~Einhorn and J.~Wudka,
  Phys.\ Rev.\  D {\bf 67}, 045004 (2003);
  M.~Creutz,
  Phys.\ Rev.\ Lett.\  {\bf 92}, 201601 (2004).




\bibitem{kharzeev2}D. Kharzeev, L. D. McLerran, H. J. Warringa
Nucl.Phys.A {\bf 803}, 227-253 (2008).

\bibitem{kharzeev3}D. Kharzeev, A. Zhitnitsky
Nucl.Phys.A {\bf 797}, 67 (2007).

\bibitem{kharzeev4}D. Kharzeev e R.D. Pisarski,
Phys.Rev.D {\bf 61},111901,2000.

\bibitem{voloshin}S. A. Voloshin,
Phys.Rev.C {\bf 62}, 044901 (2000).

\bibitem{Fraga:2008qn}
  E.~S.~Fraga and A.~J.~Mizher,
  Phys.\ Rev.\  D {\bf 78}, 025016 (2008).
\bibitem{ABJ} S. L. Adler, Phys.Rev. {\bf 177}, 2426 (1969); J.S. Bell e R. Jackiw, Nuovo Cim. A, {\bf60}, 47


\bibitem{belavin}A.A. Belavin, A.M. Polyakov, A.S. Schwartz, Y.S. Tyupkin, Phys. Lett. {\bf 37}, 172 (1976).

\bibitem{instantonrate}G. t'Hooft, Phys.Rev.Lett. {\bf 37}, 8 (1976); G. t'Hooft, Phys.Rev. D, {\bf 14}, 3432 (1976

\bibitem{sphaleronrate}G.D. Moore, C.R.Hu e B. Muller, Phys.Rev.D {\bf 58},045001 (1998); D.Bodeker, G.D.Moore e K. Rummukainen, Phys.Rev.D {\bf 61},056003 (2000)

\bibitem{levy_gellmann}M. Gell-Mann e M. Levy, Nuovo Cim. {\bf 16}, 705 (1960)

\bibitem{sigma}  M. Levy, Nuovo Cim.{\bf 52}, 23 (1967);
R.~D.~Pisarski and F.~Wilczek,
  Phys.\ Rev.\  D {\bf 29}, 338 (1984);   C.~Rosenzweig, J.~Schechter and C.~G.~Trahern,
  Phys.\ Rev.\  D {\bf 21}, 3388 (1980);
  J.~T.~Lenaghan, D.~H.~Rischke and J.~Schaffner-Bielich,
  Phys.\ Rev.\  D {\bf 62}, 085008 (2000);
  D.~Roder, J.~Ruppert and D.~H.~Rischke,
  Phys.\ Rev.\  D {\bf 68}, 016003 (2003).


\bibitem{thooft}G. t'Hooft, Phys.Rep. {\bf 142}, 357 (1986)

\bibitem{Boer:2008ct}
  D.~Boer and J.~K.~Boomsma,
  arXiv:0806.1669 [hep-ph].




\bibitem{future}A. J. Mizher and E. S. Fraga, to appear.







\end{thebibliography}
\end{document}